\documentclass[aps,preprint,groupedaddress,prd,showkeys,secnumarabic]{revtex4}
\usepackage{graphicx}

\begin{document}

% Use the \preprint command to place your local institutional report
% number in the upper righthand corner of the title page in preprint mode.
% Multiple \preprint commands are allowed.
% Use the 'preprintnumbers' class option to override journal defaults
% to display numbers if necessary
%\preprint{}

%Title of paper
\title{A novel choice of the graphene unit vectors,\\
useful in zone-folding computations}

% repeat the \author .. \affiliation  etc. as needed
% \email, \thanks, \homepage, \altaffiliation all apply to the current
% author. Explanatory text should go in the []'s, actual e-mail
% address or url should go in the {}'s for \email and \homepage.
% Please use the appropriate macro foreach each type of information

% \affiliation command applies to all authors since the last
% \affiliation command. The \affiliation command should follow the
% other information
% \affiliation can be followed by \email, \homepage, \thanks as well.
\author{P.~Marconcini}
%\email[]{Your e-mail address}
%\homepage[]{Your web page}
%\thanks{}
%\altaffiliation{}
\author{M.~Macucci
\footnote{Corresponding author. Fax: +39 050 2217522. 
Email: macucci@mercurio.iet.unipi.it  \ (M.~Macucci)}}

%  N.B.
%  Necessary to show email and fax number at the bottom of the first page

%\email[]{Your e-mail address}
%\homepage[]{Your web page}
%\thanks{}
%\altaffiliation{}
\affiliation{Dipartimento di Ingegneria dell'Informazione,
Universit\`a di Pisa\\
Via Caruso 16, I-56122 Pisa, Italy}

%Collaboration name if desired (requires use of superscriptaddress
%option in \documentclass). \noaffiliation is required (may also be
%used with the \author command).
%\collaboration can be followed by \email, \homepage, \thanks as well.
%\collaboration{}
%\noaffiliation

%\date{\today}

\begin{abstract}
\noindent
The dispersion relations of carbon nanotubes are often obtained 
cross-sectioning those of graphene (zone-folding technique) in a rectangular 
region of the reciprocal space, where it is easier to fold the resulting 
relations into the nanotube Brillouin zone. We propose a particular choice 
of the unit vectors for the graphene lattice, which consists of the symmetry 
vector and the translational vector of the considered carbon nanotube.
Due to the properties of the corresponding unit vectors in the reciprocal 
space, this choice is particularly useful for understanding the relationship 
between the rectangular region where the folding procedure is most easily 
applied and the overall graphene reciprocal space.
Such a choice allows one to find, from any graphene wave vector, the equivalent
one inside the rectangular region in a computationally inexpensive way. As an 
example, we show how the use of these unit vectors makes it easy to limit the 
computation to the bands nearest to the energy maxima and minima when 
determining the nanotube dispersion relations from those of graphene with the 
zone-folding technique.
% insert abstract here
\end{abstract}

% insert suggested PACS numbers in braces on next line
%\pacs{61.46.Fg, 73.22.-f}
% insert suggested keywords - APS authors don't need to do this

%  N.B.
%  It is necessary to add showkeys to \documentclass to make keywords appear

%\maketitle must follow title, authors, abstract, \pacs, and \keywords
\maketitle
\section{Introduction}
\begin{figure}
\includegraphics[clip,angle=0,width=14cm]{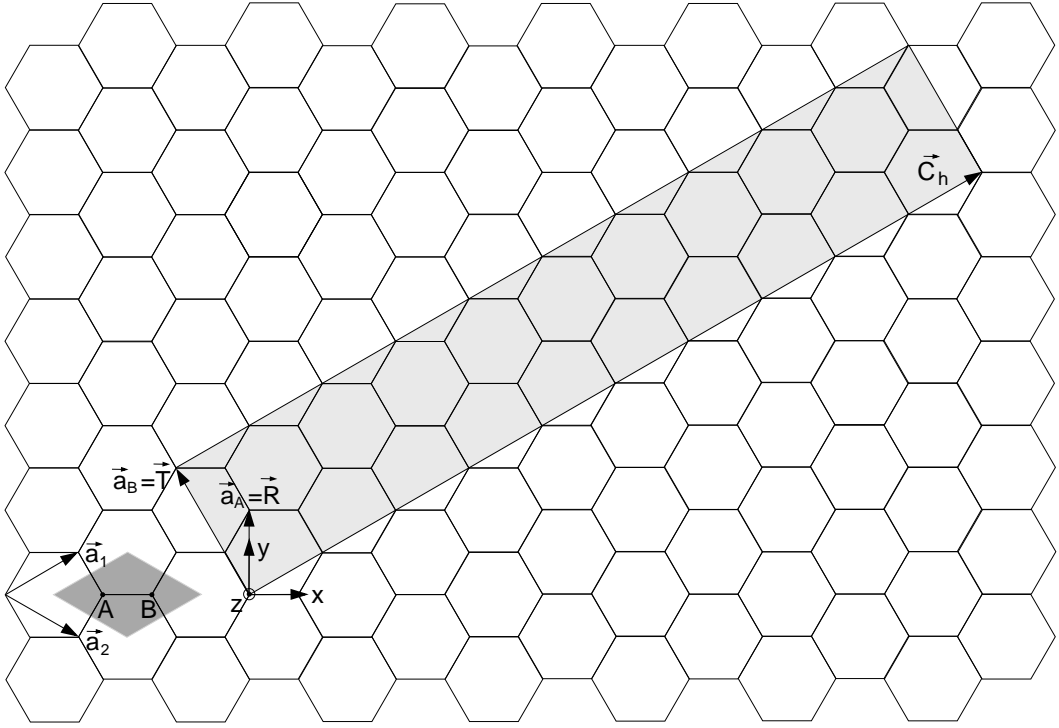}
\caption{The described quantities in the direct space for a $(10,0)$ nanotube.
The heavy-shaded rhombus is the graphene unit cell, while the 
light-shaded rectangle represents (once the graphene sheet has been rolled up) 
the nanotube unit cell.}
\label{direct}
\end{figure}

\begin{figure}
\includegraphics[clip,angle=0,width=13cm]{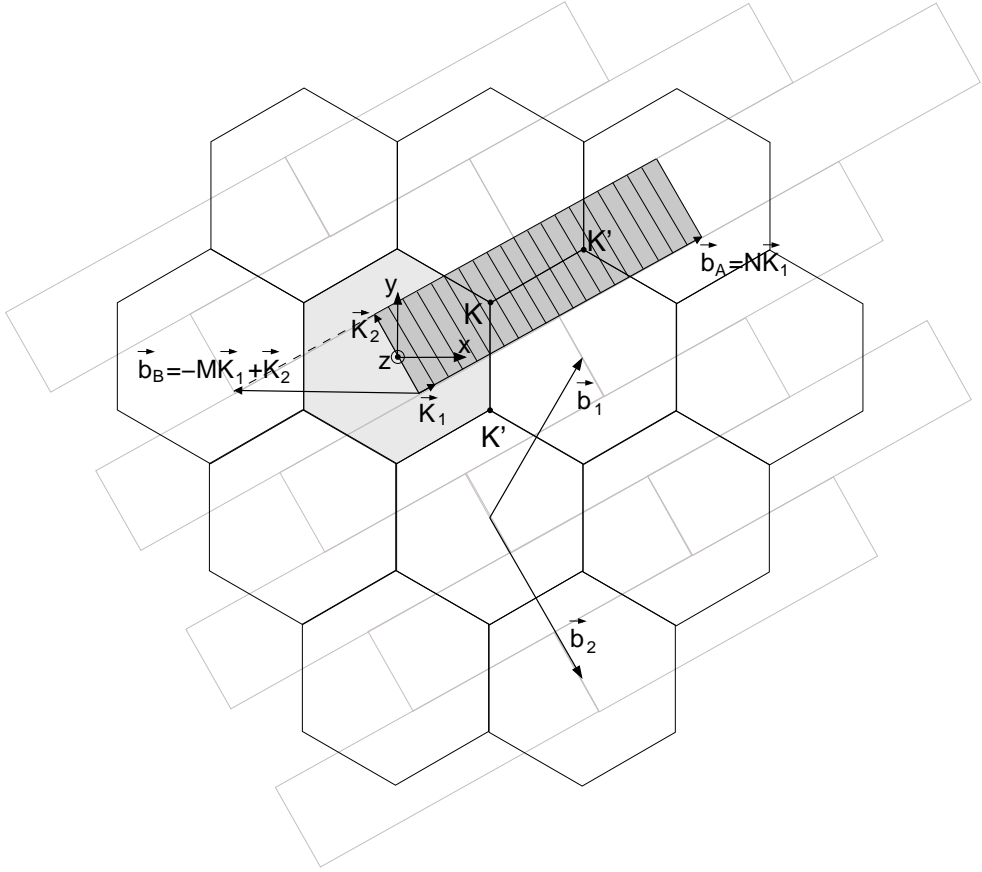}
\caption{The described quantities in the reciprocal space for a $(10,0)$ 
nanotube. The light-shaded hexagon is the graphene Brillouin zone, while 
the heavy-shaded rectangle is the rectangular area specified by the 
vectors $N \vec K_1$ and $\vec K_2$.}
\label{reciprocal}
\end{figure}

\noindent
Carbon nanotubes are cylindrical structures with diameters that are usually in 
the few nanometer range and lengths up to tens of microns. Due to their high 
mechanical strength and thermal conductivity and to their unusual electronic 
properties, carbon nanotubes constitute a very promising material for many 
applications~\cite{meyyappan,ajayan}, such as active devices, intra-chip 
interconnections, field emitters, antennas, sensors, scanning probes, 
reinforcement for composite materials, energy and hydrogen storage. In 
particular, from the electronic point of view, they can behave as metallic 
or semiconducting materials, depending on their geometrical 
properties~\cite{hamada,mintmire,saito1,saito}.\newline
A single-wall carbon nanotube can be described as a graphene sheet rolled, 
along one of its lattice vectors (the so-called chiral vector), into a 
cylindrical shape. As a consequence of the closure boundary condition along 
the chiral vector, only a subset of graphene wave vectors, located on 
parallel lines, are allowed. Therefore the dispersion relations of carbon 
nanotubes are often found cross-sectioning those of graphene along such
lines (zone-folding technique)~\cite{saito,reichbook}.
The cross-sections are usually taken in a particular rectangular region of 
the graphene reciprocal space (which can be seen as a primitive unit cell 
of the graphene reciprocal lattice).\newline
Here we introduce an unusual choice of the unit vectors in the graphene 
direct and reciprocal space, which, as a result of a direct geometrical 
relation with such a rectangle, makes it clearer how the overall reciprocal 
space can be obtained by replicating the rectangle.\newline
This allows a more complete understanding of the results of the zone-folding 
method, explaining, for example, how the periodicity of the nanotube energy 
bands arises from the computational procedure.\newline
We also show how such unit vectors make it easy to find, from any point of 
interest of the graphene reciprocal space, the equivalent wave vector located 
inside the above-mentioned rectangular region. 
At the end of the present communication, we apply this procedure to the 
graphene degeneration points, in order to compute just the nanotube bands 
nearest to the energy maxima and minima.\newline 
In Figs.~\ref{direct} and \ref{reciprocal} we show the 
graphene lattice in the direct and reciprocal space, respectively, and the 
$(\hat x,\hat y)$ reference frame that we have used in the following.
The graphene lattice structure in the real space can be seen as the 
replication of the graphene rhomboidal unit cell shown in Fig.~\ref{direct}
(containing two inequivalent carbon atoms $A$ and $B$) through linear
combinations with integer coefficients of the lattice unit vectors
$\vec a_1=(\sqrt{3}a/2) \hat x+(a/2)\hat y $ and
$\vec a_2=(\sqrt{3}a/2) \hat x-(a/2)\hat y $. Correspondingly, the unit
vectors of the graphene reciprocal lattice are 
$\vec b_1=(2 \pi /(\sqrt{3} a)) \hat x+(2 \pi / a)\hat y $
and $\vec b_2=(2 \pi /(\sqrt{3} a)) \hat x-(2 \pi / a)\hat y $, which are
reported in Fig.~\ref{reciprocal}, along with the graphene hexagonal Brillouin 
zone.\newline
An $(n,m)$ carbon nanotube is obtained rolling up a graphene sheet along its
chiral vector $\vec C_h=n \vec a_1+m \vec a_2$; the circumference of the 
nanotube is consequently equal to the length of this vector: 
$L=|\vec C_h|=a \sqrt{n^2+m^2+nm}$.\newline 
If we define $d_R$ as the greatest common divisor of $2m+n$ and $2n+m$, 
we have that the lattice unit vector of the nanotube (which represents 
a 1D lattice) is the so-called translational vector
$\vec T=t_1 \vec a_1+t_2 \vec a_2$ of the unrolled graphene sheet, 
parallel to the nanotube axis and orthogonal to $\vec C_h$, where 
$t_1$ and $t_2$ are relatively prime integer numbers given by 
$t_1=(2m+n)/d_R$ and $t_2=-(2n+m)/d_R$.
Therefore, the rectangle having as edges the chiral vector $\vec C_h$ and the 
translational vector $\vec T$ represents the unit cell of the nanotube, which 
repeats identically along the nanotube axis with a lattice unit vector 
$\vec T$, the length of which is equal to $T=|\vec T|=\sqrt{3}L / d_R$. The 
number of graphene unit cells inside the nanotube unit cell is equal to 
$N=2 L^2/(a^2 d_R)$.\newline
The coordinates of all the $N$ points identifying the graphene unit cells
inside the rectangular region representing the nanotube unit cell in the 
unrolled graphene sheet are defined (apart from translations by an integer 
number of $\vec C_h$ and $\vec T$ vectors) by integer multiples of the 
so-called symmetry vector $\vec R=p \vec a_1+q \vec a_2 $, where $p$ and $q$ 
are two relatively prime integer numbers, univocally determined by the 
two relations: $t_1 q-t_2 p=1$ and $0 < M \le N $ (where we define the 
quantity $M=mp-nq$). In particular, we have that 
$N \vec R=\vec C_h+M \vec T$~\cite{saito}.\newline
In Fig.~\ref{direct} we show all of these vectors in the direct space for the
nanotube $(10,0)$ (for which $\vec C_h=10 \vec a_1$, $L=10a$, $d_R=10$, 
$\vec T=\vec a_1-2\vec a_2$, $T=\sqrt{3}a$, $N=20$, $\vec R=\vec a_1-\vec a_2$ 
and $M=10$).
%\hspace{-10pt}
\par\noindent
Since in the direct space the nanotube is a one-dimensional lattice with a 
unit vector $\vec T$ and with a $T$ wide unit cell along the nanotube axis, 
in the reciprocal space its unit vector is equal to 
$\vec K_2=(2 \pi / T) \hat T=(m \vec b_1-n \vec b_2)/N$ and its Brillouin 
zone is represented by the values of the wave vector $k$ (along the nanotube 
axis) which satisfy the inequality $-\pi / T < k \le \pi / T$.\newline
Many physical properties of the nanotubes, such as the energy dispersion 
relations, can be found from the corresponding quantities of graphene using the
zone-folding technique~\cite{saito,reichbook}. Indeed, as a consequence of the 
closure of the graphene sheet to form the carbon nanotube, we have to enforce 
that the graphene electron wave function $e^{i \vec k \cdot \vec r} u(\vec k)$ 
(where $u(\vec k)$ is a Bloch lattice function) has identical values in 
any pair of points $\vec r$ and $\vec r+\vec C_h$. To satisfy the resulting 
relation $e^{i \vec k \cdot \vec C_h}=1$, the component along $\hat C_h$ 
of the wave vector has to be equal to an integer multiple of the vector 
$\vec K_1=(2 \pi / L)\hat C_h=(-t_2 \vec b_1+t_1 \vec b_2)/N$.
If we cross-section the graphene dispersion relations in correspondence of
the parallel lines (separated by a distance $|\vec K_1|=2 \pi / L$) containing
the allowed graphene wave vectors and we fold such sections
into the nanotube Brillouin zone, we find the relations for the 
carbon nanotube. This procedure is applied to a region of the graphene
reciprocal space containing all and only the inequivalent graphene 
wave vectors. In particular, the rectangle having as edges the vectors 
$N \vec K_1$ and $\vec K_2$ has these properties (as we have demonstrated in 
the Supplementary Information) and corresponds to the region that is usually 
implicitly chosen to apply the zone-folding method, because here, considering 
only the allowed graphene wave vectors, we obtain $N$ segments with a width 
equal to $2 \pi / T$, that can be folded into the Brillouin zone of the 
nanotube by simply taking the component along $\hat T$ of each graphene 
wave vector (such a component becomes the nanotube wave vector).\newline
In Fig.~\ref{reciprocal} we show the quantities in the reciprocal space 
for the nanotube $(10,0)$ (for which 
$\vec K_1=(2 \pi /(10a))\hat C_h=(2 \vec b_1+\vec b_2)/20$ and 
$\vec K_2=(2 \pi /(\sqrt{3}a)) \hat T=-\vec b_2/2$).
\par\noindent
We propose an alternative choice of the graphene unit vectors
that allows to clarify the relation between the previously described 
rectangular region specified by the vectors $N \vec K_1$ and $\vec K_2$ and
the overall graphene reciprocal space; in particular it makes it easy to
find, for any given wave vector, the equivalent wave vector belonging to 
such a rectangular area. As we shall see, this can be very useful when 
we apply the zone-folding method cross-sectioning this rectangle.
\par\noindent
\section{Alternative choice of the graphene unit vectors}
\noindent
Let us consider the graphene sheet forming, once rolled up, a carbon nanotube 
with chiral vector $\vec C_h$, translational vector $\vec T$ and symmetry 
vector $\vec R$. We propose, as an alternative choice of the graphene 
unit vectors in the direct space, exactly the pair of vectors 
$\vec a_A \equiv \vec R$ and $\vec a_B \equiv \vec T$:
\begin{eqnarray}
\vec a_A \!&=&\! \vec R=p \vec a_1+q \vec a_2=
\frac{a}{2}\sqrt{3}(p+q)\hat x+\frac{a}{2}(p-q)\hat y \\
\nonumber\\[-10pt]
\vec a_B \!&=&\! \vec T=t_1 \vec a_1+t_2 \vec a_2=
\frac{a}{2}\sqrt{3}(t_1+t_2)\hat x+\frac{a}{2}(t_1-t_2)\hat y \quad .
\end{eqnarray}
To verify that these two vectors can actually be used as graphene unit vectors 
in the direct space, we have to demonstrate that their linear combinations
with integer coefficients yield all and only the lattice vectors of
such a space, i.e. the vectors that are also linear combinations with integer
coefficients of $\vec a_1$ and $\vec a_2$.\newline
Since both $\vec R$ and $\vec T$ are linear combinations with integer
coefficients of $\vec a_1$ and $\vec a_2$, every linear
combination with integer coefficients of $\vec R$ and $\vec T$ is also
a linear combination with integer coefficients of $\vec a_1$ and 
$\vec a_2$.\newline
On the other hand, in order to demonstrate that every linear combination 
with integer coefficients of $\vec a_1$ and $\vec a_2$ is also a linear 
combination with integer coefficients of $\vec R$ and $\vec T$, it is 
useful to consider the linear system consisting of the two following known 
relations:
\begin{equation}
\left\{ \begin{array}{l}
p \vec a_1 +q \vec a_2= \vec R \\
t_1 \vec a_1 +t_2 \vec a_2= \vec T \\
\end{array} \right. \quad.
\end{equation}
Solving for $\vec a_1$ and $\vec a_2$, we find that
\begin{equation}
\left\{ \begin{array}{l}
(-t_2 p+q t_1) \vec a_1= -t_2 \vec R +q \vec T \\
(t_1 q-p t_2) \vec a_2= t_1 \vec R -p \vec T \\
\end{array} \right. \quad \textrm{or equivalently} \quad
\left\{ \begin{array}{l}
\vec a_1= -t_2 \vec R +q \vec T \\
\vec a_2= t_1 \vec R -p \vec T \\
\end{array} \right.
\end{equation}
(recalling that $t_1 q-t_2 p=1$), which means that $\vec a_1$ and $\vec a_2$
are linear combinations with integer coefficients of $\vec R$ and $\vec T$.
Consequently, also every linear combination with integer coefficients of 
$\vec a_1$ and $\vec a_2$ is a linear combination with integer coefficients 
of $\vec R$ and $\vec T$.\newline
Using the coordinates of $\vec R$ and $\vec T$ in the $(\hat x,\hat y)$ 
reference frame of Fig.~\ref{direct}, and introducing a unit vector $\hat z$
that is orthogonal to the plane $\hat x,\hat y$ of the graphene sheet and 
forms a right-hand reference frame $(\hat x, \hat y, \hat z)$ with $\hat x$ 
and $\hat y$, we have that 
\begin{eqnarray}
\vec T \times \hat z \!&=&\! -\frac{a}{2} (t_2-t_1)\hat x-
\frac{a}{2}\sqrt{3} (t_2+t_1)\hat y \\
\nonumber\\[-10pt]
\hat z \times \vec R \!&=&\! \frac{a}{2} (q-p)\hat x+
\frac{a}{2}\sqrt{3} (q+p)\hat y \\
\nonumber\\[-10pt]
\vec R \cdot \left(\vec T \times \hat z \right) \!&=&\!
\frac{a^2 \sqrt{3}}{2} (t_1 q-t_2 p)=\frac{a^2 \sqrt{3}}{2} \quad ,
\end{eqnarray}
where we have used again the relation $t_1 q-t_2 p=1$.
Therefore, the corresponding unit vectors of graphene in the reciprocal space 
are (using the well-known relations between the unit vectors in the direct and 
in the reciprocal space~\cite{ashcroft}):
\begin{eqnarray}
\vec b_A \!&=&\!
2 \pi \frac{\vec a_B \times \hat z}{\vec a_A \cdot (\vec a_B \times \hat z)}=
2 \pi \frac{\vec T \times \hat z}{\vec R \cdot (\vec T \times \hat z)}=
-\frac{2 \pi}{a \sqrt{3}} (t_2-t_1)\hat x-\frac{2 \pi}{a}(t_2+t_1)\hat y=
\nonumber\\
\nonumber\\[-10pt]
\!&=&\! -t_2 \left( \frac{2 \pi}{\sqrt{3} a} \hat x+
\frac{2 \pi}{a}\hat y \right) +
t_1 \left( \frac{2 \pi}{\sqrt{3} a} \hat x-\frac{2 \pi}{a}\hat y \right) =
-t_2 \vec b_1+t_1 \vec b_2= \nonumber\\
\nonumber\\[-10pt]
\!&=&\! -t_2 (n \vec K_1+t_1 \vec K_2)+t_1 (m \vec K_1+t_2 \vec K_2)=
(-t_2 n+t_1 m) \vec K_1= \nonumber\\
\nonumber\\[-10pt]
\!&=&\! N \vec K_1 =N \frac{2 \pi}{L} \hat C_h
\end{eqnarray}
and
\begin{eqnarray}
\vec b_B \!&=&\!
2 \pi \frac{\hat z \times \vec a_A}{\vec a_A \cdot (\vec a_B \times \hat z)}=
2 \pi \frac{\hat z \times \vec R}{\vec R \cdot (\vec T \times \hat z)}=
\frac{2 \pi}{a \sqrt{3}} (q-p)\hat x+\frac{2 \pi}{a}(q+p)\hat y=
\nonumber\\
\nonumber\\[-10pt]
\!&=&\! q \left( \frac{2 \pi}{\sqrt{3} a} \hat x+
\frac{2 \pi}{a}\hat y \right) -
p \left( \frac{2 \pi}{\sqrt{3} a} \hat x-\frac{2 \pi}{a}\hat y \right)=
q \vec b_1-p \vec b_2= \nonumber\\
\nonumber\\[-10pt]
\!&=&\! q (n \vec K_1+t_1 \vec K_2)-p (m \vec K_1+t_2 \vec K_2)=
(qn-pm) \vec K_1+(q t_1-p t_2) \vec K_2= \nonumber\\
\nonumber\\[-10pt]
\!&=&\! -M \vec K_1+\vec K_2 = 
-M \frac{2 \pi}{L} \hat C_h + \frac{2 \pi}{T} \hat T
\quad ,
\end{eqnarray}
which are linear combinations with integer coefficients of the vectors 
$\vec K_1$ and $\vec K_2$ and have components along $\hat C_h$ and 
$\hat T$: $b_{A_C}=N (2 \pi / L)$~, \ $b_{A_T}=0$~, \
$b_{B_C}=-M (2 \pi / L)$~, \ $b_{B_T}=2 \pi / T $~.
To find the previous results we have used, besides the fact that 
$t_1 q-t_2 p=1$ and $M=mp-nq$, the relations: $-t_2 n+t_1 m=N$,
$\vec b_1=n \vec K_1 +t_1 \vec K_2$ and
$\vec b_2=m \vec K_1 +t_2 \vec K_2$.\newline 
Indeed, using the relations listed in the Introduction, we obtain that
\begin{equation}
m t_1 -n t_2=m \frac{2m+n}{d_R}+n \frac{2n+m}{d_R}=
\frac{2}{d_R} (m^2+nm+n^2)=
\frac{2}{d_R} \left(\frac{L}{a} \right)^2 =N \quad.
\end{equation}
The relations between the vectors $\vec b_1$ and $\vec b_2$ and the 
vectors $\vec K_1$ and $\vec K_1$ can instead be found starting
from the identities:
\begin{equation}
\left\{ \begin{array}{l}
\displaystyle \vec K_1=\frac{1}{N} (-t_2 \vec b_1+t_1 \vec b_2)\\
\\[-10pt]
\displaystyle \vec K_2=\frac{1}{N} (m \vec b_1-n \vec b_2)\\
\end{array} \right.
\end{equation}
and solving for $\vec b_1$ and $\vec b_2$ this system of two equations. 
We find that
\begin{equation}
\left\{ \begin{array}{l}
(m t_1 -n t_2) \vec b_1=N (n \vec K_1 +t_1 \vec K_2) \\
(m t_1 -n t_2) \vec b_2=N (m \vec K_1 +t_2 \vec K_2) \\
\end{array} \right.
\end{equation}
and thus (using the fact that $m t_1 -n t_2=N$) that
\begin{equation}
\left\{ \begin{array}{l}
\displaystyle \vec b_1=n \vec K_1 +t_1 \vec K_2=
n \frac{2 \pi}{L} \hat C_h +t_1 \frac{2 \pi}{T} \hat T \\
\\[-10pt]
\displaystyle \vec b_2=m \vec K_1 +t_2 \vec K_2=
m \frac{2 \pi}{L} \hat C_h +t_2 \frac{2 \pi}{T} \hat T \\
\end{array} \right. \quad .
\end{equation}
From this result, it is also apparent that the components of $\vec b_1$ and 
$\vec b_2$ along $\hat C_h$ and $\hat T$ are:
$b_{1_C}=n (2 \pi / L)$~, \  
$b_{1_T}=t_1 (2 \pi / T)$~, \ 
$b_{2_C}=m (2 \pi / L)$~, \
$b_{2_T}=t_2 (2 \pi / T)$~.
\par\noindent
\section{Applications}
\begin{figure}
\includegraphics[clip,angle=0,width=9.5cm]{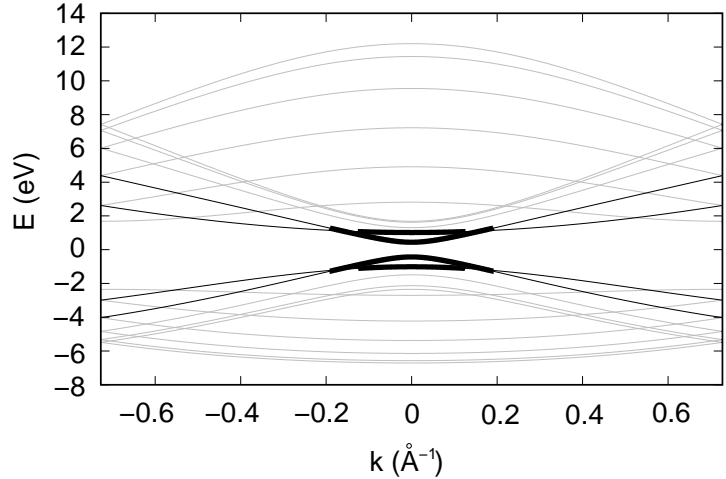}
\caption{Results of the proposed optimization of the tight-binding method 
in the case of a $(10,0)$ nanotube. The thin black lines are the (doubly 
degenerate) bands obtained cross-sectioning the graphene dispersion 
relations along the segments closest to the two graphene degeneration 
points inside the considered rectangle and the thick black lines are the 
parts of these bands obtained taking the cross-sections only in the circular 
regions (with radius equal to $(5/6)\,|\vec K_1|$) centered around the 
two graphene degeneration points. The gray curves represent the nanotube 
dispersion relations that with our procedure we do not need to compute, if
we are not interested in them.}
\label{bands}
\end{figure}

\noindent
This choice of unit vectors helps us to understand the relation
between the rectangular region having as edges the vectors $N \vec K_1$ 
and $\vec K_2$ and the overall graphene reciprocal space. 
Indeed, such a rectangular region contains all and only the inequivalent 
graphene wave vectors and can therefore be considered as a primitive 
unit cell of the graphene reciprocal lattice. In particular, 
considering as unit vectors of the graphene reciprocal space $\vec b_A$ and 
$\vec b_B$ (which have a clear geometrical relation with the considered 
region), we have that the overall reciprocal space can be spanned translating 
the rectangular region by vectors that are linear combinations, with 
integer coefficients, of $\vec b_A$ (which is exactly equal to $N \vec K_1$,
the base of the rectangle) and $\vec b_B$ (which has a component along 
$\hat T$ equal to $|\vec K_2|$, the height of the rectangle, and a component 
along $\hat C_h$ equal to $-M |\vec K_1|$, i.e. an integer number of times 
the distance $|\vec K_1|$ between the segments along which we take the $N$ 
cross-sections inside the rectangle in the zone-folding method). Therefore,
the overall graphene reciprocal space is spanned by rows (parallel to 
$\hat C_h$) of equivalent rectangles, with each row shifted along 
$\hat C_h$ by $-M \vec K_1$ with respect to the adjacent one (as we show
in gray in Fig.~\ref{reciprocal} for the particular case of a (10,0) 
nanotube, where $M=10$, $N=20$ and therefore $N=2M$).\newline
This clarifies the result obtained by cross-sectioning the graphene dispersion 
relations along the $N$ parallel lines (separated by a distance $|\vec K_1|$) 
to which the $N$ parallel segments used in the zone-folding method belong.
Since the parallel rows of rectangles spanning the graphene reciprocal space 
have a nonzero relative shift along $\hat C_h$ (and therefore the generic 
graphene wave vectors $\vec k$ and $\vec k+\vec K_2$ are not equivalent),
the relation obtained from each single cross-section in general is not
periodic with period equal to $|\vec K_2|$ (the width of the nanotube 
Brillouin zone). Nevertheless, since the relative shift along $\hat C_h$ 
between rows of rectangles is an integer multiple of the distance 
$|\vec K_1|$ between the $N$ parallel lines, we find, starting from a 
wave vector $\vec k$ on one of the $N$ lines and moving by $|\vec K_2|$ 
along $\hat T$ and by a proper multiple of $|\vec K_1|$ along $\vec C_h$, 
a wave vector equivalent to $\vec k$ on another of the $N$ lines of allowed 
wave vectors. Therefore the overall set of relations, obtained drawing all 
the $N$ cross-sections on the same one-dimensional domain, is indeed 
periodic with period $|\vec K_2|$. This is in agreement with the fact 
that the resulting relations are the nanotube dispersion relations, 
that have to be periodic with a period equal to the width of the nanotube 
Brillouin zone.\newline
The proposed alternative choice of graphene unit vectors is particularly  
useful for the determination, for any given wave vector, of the equivalent 
wave vector within the previously discussed rectangular region of 
the graphene reciprocal space.\newline
Indeed, given a graphene wave vector $\vec k$, if we use $\vec b_A$ and
$\vec b_B$ as unit vectors in the reciprocal space, all the wave
vectors equivalent to $\vec k$ can be written as 
$\vec k^{eq}=\vec k+\alpha \vec b_A+\beta \vec b_B$, 
with $\alpha$ and $\beta$ integer numbers. Thus the corresponding components 
$k^{eq}_C$ and $k^{eq}_T$ along $\hat C_h$ and $\hat T$, respectively, are:
\begin{equation}
\left\{ \begin{array}{l}
\displaystyle k^{eq}_C= k_C+\alpha b_{A_C}+\beta b_{B_C}=
k_C+\alpha N \frac{2 \pi}{L}-\beta M \frac{2 \pi}{L} \\
\\[-10pt]
\displaystyle k^{eq}_T= k_T+\alpha b_{A_T}+\beta b_{B_T}=
k_T+\beta \frac{2 \pi}{T}
\end{array} \right. \quad .
\end{equation}
Since we want to find the particular $\vec k^{eq}$ belonging to the 
rectangular region, such components have to satisfy the following relations:
\begin{equation}
\left\{ \begin{array}{l}
\displaystyle  0 \le k^{eq}_C < N \frac{2 \pi}{L} \\
\\[-10pt]
\displaystyle  -\frac{\pi}{T} < k^{eq}_T \le \frac{\pi}{T}
\end{array} \right.  \quad \textrm{or equivalently} \quad
\left\{ \begin{array}{l}
\displaystyle  0 \le k^{eq}_C < N \frac{2 \pi}{L} \\
\\[-10pt]
\displaystyle  0 < k^{eq}_T+\frac{\pi}{T} \le \frac{2 \pi}{T}
\end{array} \right. \quad .
\end{equation}
Substituting the expressions of $k^{eq}_C$ and $k^{eq}_T$ into these
inequalities, we find:
\begin{equation} \label{unsy1}
\left\{ \begin{array}{l}
\displaystyle 
0 \le \left( k_C-\beta M \frac{2 \pi}{L}\right)+\alpha N \frac{2 \pi}{L} 
< N \frac{2 \pi}{L} \\
\\[-10pt]
\displaystyle 
0 < \left( k_T+\frac{\pi}{T}\right)+\beta \frac{2 \pi}{T} 
\le \frac{2 \pi}{T}
\end{array} \right.
\end{equation}
and thus
\begin{equation} \label{unsy2}
\left\{ \begin{array}{l}
\displaystyle 
-\alpha \left( N \frac{2 \pi}{L} \right) \le k_C-\beta M \frac{2 \pi}{L}
< (-\alpha +1) \left( N \frac{2 \pi}{L} \right) \\
\\[-10pt]
\displaystyle 
-\beta \left( \frac{2 \pi}{T} \right) < k_T+\frac{\pi}{T}
\le (-\beta+1) \left(\frac{2 \pi}{T} \right)
\end{array} \right.
\end{equation}
or equivalently
\begin{equation} \label{unsy3}
\left\{ \begin{array}{l}
\displaystyle 
-\alpha \le \frac{k_C-\beta M (2 \pi / L)}{N (2 \pi / L)}
< -\alpha +1 \\
\\[-10pt]
\displaystyle 
-\beta < \frac{k_T+(\pi / T)}{2 \pi / T}
\le -\beta+1
\end{array} \right. \quad .
\end{equation}
The values of $\alpha$ and $\beta$, and consequently the vector 
$\vec k^{eq}$, can be easily found using the fact that the second inequality 
contains only $\beta$. Indeed, from the second inequality we find that
(using the ceiling and floor functions):
\begin{equation}
\beta=\left\{ \begin{array}{ll}
\displaystyle 
-\left\lceil\, \frac{k_T +(\pi / T)}{2 \pi / T} \,\right\rceil+1
& \textrm{if \ $\displaystyle k_T+\frac{\pi}{T} \ge 0$ } \\
\\[-10pt]
\displaystyle
\left\lfloor \frac{|k_T +(\pi / T)|}{2 \pi / T} \right\rfloor +1
& \textrm{if \ $\displaystyle k_T+\frac{\pi}{T} < 0$ }
\end{array} \right. \quad .
\end{equation}
Once the value of $\beta$ is found, the quantity $k_C-\beta M (2 \pi / L)$ 
in the first inequality is known and thus from first inequality we obtain 
that:
\begin{equation}
\alpha=\left\{ \begin{array}{ll}
\displaystyle 
-\left\lfloor \frac{k_C -\beta M (2 \pi / L)}{N (2 \pi / L)} \right\rfloor
& \textrm{if \ $\displaystyle k_C-\beta M \frac{2 \pi}{L} \ge 0$ } \\
\\[-10pt]
\displaystyle
\left\lceil\, \frac{|k_C -\beta M (2 \pi / L)|}{N (2 \pi / L)} \,\right\rceil
& \textrm{if \ $\displaystyle k_C-\beta M \frac{2 \pi}{L} < 0$ }
\end{array} \right. \quad .
\end{equation}
Clearly the second inequality of the systems (\ref{unsy1})--(\ref{unsy3}) does 
not contain $\alpha$ only because, with our particular choice of unit vectors, 
$\vec b_A$ has a zero component along $\hat T$ (i.e. $b_{A_T}=0$).\newline 
In the following we show an application of this procedure for the optimization
of the zone-folding computation of the nanotube energy bands.\newline 
As we have described, the nanotube dispersion relations can be obtained
cross-sectioning the bands of graphene (computed for example 
with the tight-binding method), in the rectangular region of the reciprocal 
space specified by the vectors $N \vec K_1$ and $\vec K_2$, along the $N$ 
equidistant segments, parallel to $\hat T$, containing all the wave vectors 
of the region with component along $\hat C_h$ multiple of 
$|\vec K_1|=2 \pi / L$. In this way, cross-sectioning the
two energy bands (bonding and anti-bonding) of graphene, we obtain $2 N$
dispersion relations that, once folded into the nanotube Brillouin zone
(which coincides with the first segment, along the nanotube axis),
form the nanotube energy bands.\newline
Among these $2 N$ energy bands, the most interesting ones are the lowest
conduction bands and the highest valence bands, because these are the regions 
where the charge carriers localize. These bands are obtained cross-sectioning 
the graphene dispersion relations near the particular graphene wave vectors
\begin{eqnarray}
\label{K}
\vec K \!&=&\! \frac{2 \pi}{\sqrt{3} a}\hat x+\frac{2 \pi}{3 a}\hat y=
\frac{1}{3}(2 \vec b_1+\vec b_2)=
\frac{1}{3}(2 b_{1_C}+b_{2_C})\hat C_h+
\frac{1}{3}(2 b_{1_T}+b_{2_T})\hat T \quad , \\
\nonumber\\[-10pt]
\label{K1}
\vec K' \!&=&\! \frac{2 \pi}{\sqrt{3} a}\hat x-\frac{2 \pi}{3 a}\hat y=
\frac{1}{3}(\vec b_1+2 \vec b_2)=
\frac{1}{3}(b_{1_C}+2 b_{2_C})\hat C_h+
\frac{1}{3}(b_{1_T}+2 b_{2_T})\hat T
\end{eqnarray}
and their equivalent wave vectors, where the graphene energy bands have their 
maxima and minima (and in particular are degenerate). Therefore we need
to find the wave vectors equivalent to $\vec K$ and $\vec K'$ inside the 
rectangular region where we take the cross-sections of the graphene dispersion 
relations (we shall find just one wave vector equivalent to $\vec K$ and just 
one equivalent to $\vec K'$). This is done applying the previously described 
procedure to $\vec K$ and $\vec K'$, whose components along $\hat C_h$ and 
$\hat T$ are given by Eqs.~(\ref{K})-(\ref{K1}). In particular, we 
know~\cite{saito} that the components along $\hat T$ of the graphene wave 
vectors equivalent to $\vec K$ and $\vec K'$ and belonging to the rectangle 
can only assume the values $0$ or $\pm 2 \pi / (3 T)$ and therefore are well 
inside the considered region, away from the boundaries. Once we have found 
these wave vectors, we can cross-section the graphene dispersion relations 
just along the segments in their proximity, if we want only the mentioned
most relevant energy bands. This leads to a nonnegligible reduction of the 
computational effort.\newline 
For example, in Fig.~\ref{bands} we show the results obtained for a $(10,0)$ 
carbon nanotube (for which, as we have seen, $N=20$). The
graphene dispersion relations have been computed with the tight-binding 
method, in which we have considered only the $2 p_z$ orbital for each 
atom and have included the effect on each atom of up to the third-nearest 
neighbors. In this computation we have used, for the tight-binding 
parameters, the values found in \cite{reich} fitting in the optical 
range the results of an {\it ab initio} calculation performed with the 
SIESTA code: $\varepsilon_{2p}=-2.03$~eV, $\gamma_0=-2.79$~eV, 
$s_0=0.30$, $\gamma_1=-0.68$~eV, $s_1=0.046$, $\gamma_2=-0.30$~eV and 
$s_2=0.039$ (with the notation used in \cite{reich}). We have considered 
10001 wave vector values on each of the $N$ parallel segments along which the 
graphene energy bands are cross-sectioned (in order to keep our code as general
as possible, we did not exploit the particular simmetry properties of the 
achiral $(10,0)$ nanotube). The curves shown in the figure represent all the 
$2N=40$ (partially degenerate) bands (20 bonding bands and 20 anti-bonding 
bands) of the $(10,0)$ nanotube. In particular, with the thin black lines we 
represent the 8 bands (degenerate in pairs) obtained cross-sectioning the 
graphene dispersion relations along the segments closest to the two graphene 
degeneration points inside the considered rectangle, and with the thick black 
lines we report the portions of these bands obtained taking the cross-sections 
only in the circular regions (with radius equal to $(5/6)\,|\vec K_1|$) 
centered around the two graphene degeneration points. The data represented 
with the (thin and thick) black lines have been obtained with the previously 
described method.\newline
In our simulations we have found that, on a Pentium~4 at 2.4~GHz, the 
time spent  to find all the bands of the $(10,0)$ nanotube is 200~ms, 
five times greater than the time (40~ms) taken by the modified 
version of the program, which computes only the bands closest to the two 
graphene degeneration points: the computational time substantially scales 
proportionally to the number of computed bands. Computing only the parts of 
such bands closest to the graphene degeneration points we have a further 
speed-up: in this case the time spent becomes 10~ms. The proposed improvement 
becomes actually useful in the situations in which many calculations of this 
type need to be performed, leading to significant computational times.
\par\noindent
Incidentally, we note that an alternative method to calculate only the 
most relevant nanotube energy bands could consist in taking the cross-sections 
of the graphene dispersion relations (along the parallel lines corresponding 
to the allowed wave vectors) in the hexagonal Brillouin zone of graphene 
(which evidently contains all and only the inequivalent graphene wave vectors),
instead of inside the considered rectangular region. In this case the 
positions of the maximum and minimum points are well known ($\vec K$ and 
$\vec K'$ are at the vertices of the hexagon) and thus the regions of interest 
are clearly located. Following this method, however, in order to fold the 
computed cross-sections into the nanotube Brillouin zone (the segment of the 
$\hat T$ axis characterized by $-\pi / T < k_T \le \pi / T$), it is not 
sufficient to consider the $k_T$ component along $\hat T$ of the graphene
wave vector (the absolute value of this component can be greater than 
$\pi / T$), but we also need to find the nanotube wave vector equivalent to 
$k_T$ inside the nanotube Brillouin zone. Moreover, since the graphene 
degeneration points are on the boundary of the sectioned hexagonal region, 
in order to obtain the nanotube bands around their minima and maxima we have 
to properly join the results computed cross-sectioning the graphene dispersion 
relations near the six vertices (each of which gives only parts of the desired 
bands); this strongly increases the algorithmic complexity of the involved 
computations.
\par\noindent
\section{Conclusion}
\noindent
We have proposed an alternative choice of unit vectors for the graphene sheet 
which, once rolled up, forms a carbon nanotube. These vectors, which depend on 
the considered nanotube, are closely related to the rectangular region of the 
graphene reciprocal space where the zone-folding method is most easily applied,
and allow us to better understand the relation between this rectangular area 
and the overall reciprocal space. In particular, we have shown that our choice 
of unit vectors can be exploited to find, from any graphene wave vector, the 
equivalent wave vector inside the rectangular region and can therefore be 
useful whenever the zone-folding technique is applied to obtain the physical 
properties of the carbon nanotube from those of graphene. As an example, we 
have presented an application to the optimization of the tight-binding 
calculation of the carbon nanotube energy dispersion relations, with a 
significant reduction of computational times.
\par\noindent
\begin{acknowledgments}
\par\noindent
We acknowledge Dr.~Michele Pagano for useful discussions.\newline
This work has been supported by the Italian Ministry of Education, University 
and Research (MIUR) through the FIRB project ``Nanotechnologies and
Nanodevices for the Information Society''.
\end{acknowledgments}
\par\noindent
\section*{Supplementary Information}
%
% N.B.
% the * allows to avoid the number
%
\par\noindent
In the following, we provide a demonstration of the fact that the rectangular 
region of the graphene reciprocal space defined by the vectors $N \vec K_1$ 
and $\vec K_2$ contains all and only the inequivalent graphene wave vectors 
and can therefore be considered as a primitive unit cell of the graphene 
reciprocal lattice.\newline
This rectangular region has an area
\begin{equation}
N \frac{2 \pi}{L} \cdot \frac{2 \pi}{T}=
\frac{2 L^2}{a^2 d_R} \frac{2 \pi}{L} 2 \pi \frac{d_R}{\sqrt{3} L}=
\frac{8 \pi^2}{\sqrt{3} a^2} \quad ,
\end{equation}
that is equal to $|\vec b_1 \times \vec b_2|$ and thus to the area of the
graphene Brillouin zone. Thus our assertion is automatically verified if we
demonstrate that such a region does not contain two distinct but
equivalent (i.e. differing only for a linear combination with integer
coefficients of $\vec b_1$ and $\vec b_2$) graphene wave vectors 
$\vec k=k_C \hat C_h+k_T \hat T $ and 
$\vec k'=\vec k+N_1 \vec b_1+N_2 \vec b_2=k'_C \hat C_h+k'_T \hat T $
(with $N_1$ and $N_2$ two integer numbers).\newline
Indeed, if $\vec k$ is inside the rectangular region, we have that
\begin{equation}
\left\{ \begin{array}{l}
\displaystyle 0 \le k_C < N \frac{2 \pi}{L} \\
\\[-10pt]
\displaystyle -\frac{\pi}{T}< k_T \le \frac{\pi}{T} \\
\end{array} \right. \quad \textrm{and therefore} \quad
\left\{ \begin{array}{l}
\displaystyle -N \frac{2 \pi}{L} < -k_C \le 0 \\
\\[-10pt]
\displaystyle -\frac{\pi}{T} \le -k_T < \frac{\pi}{T} \\
\end{array} \right. \quad .
\end{equation}
If also $\vec k'$ is inside the rectangular region, we have that
\begin{equation}
\left\{ \begin{array}{l}
\displaystyle 0 \le k'_C < N \frac{2 \pi}{L} \\
\\[-10pt]
\displaystyle -\frac{\pi}{T}< k'_T \le \frac{\pi}{T} \\
\end{array} \right. \quad \textrm{or equivalently} \quad
\left\{ \begin{array}{l}
\displaystyle 0 \le k_C+N_1 b_{1_C}+N_2 b_{2_C} < N \frac{2 \pi}{L} \\
\\[-10pt]
\displaystyle -\frac{\pi}{T}< k_T+N_1 b_{1_T}+N_2 b_{2_T} \le \frac{\pi}{T} \\
\end{array} \right. \quad .
\end{equation}
From these last relations, exploiting the inequalities for $-k_C$ and $-k_T$, 
we have that 
\begin{equation}
\left\{ \begin{array}{l}
\displaystyle -N \frac{2 \pi}{L} < -k_C \le N_1 b_{1_C}+N_2 b_{2_C} < 
-k_C+N \frac{2 \pi}{L} \le N \frac{2 \pi}{L} \\
\\[-10pt]
\displaystyle -\frac{2 \pi}{T} \le  -k_T-\frac{\pi}{T} < 
N_1 b_{1_T}+N_2 b_{2_T} \le -k_T+\frac{\pi}{T} < \frac{2 \pi}{T} \\
\end{array} \right. \quad .
\end{equation}
This implies that that
\begin{equation} \label{uneq}
\left\{ \begin{array}{l}
\displaystyle -N \frac{2 \pi}{L} < N_1 b_{1_C}+N_2 b_{2_C} < 
N \frac{2 \pi}{L} \\
\\[-10pt]
\displaystyle -\frac{2 \pi}{T} < N_1 b_{1_T}+N_2 b_{2_T} < 
\frac{2 \pi}{T} \\
\end{array} \right. \quad \textrm{and thus} \quad
\left\{ \begin{array}{l}
\displaystyle |N_1 b_{1_C}+N_2 b_{2_C}| < N \frac{2 \pi}{L} \\
\\[-10pt]
\displaystyle |N_1 b_{1_T}+N_2 b_{2_T}| < \frac{2 \pi}{T} \\
\end{array} \right. \quad .
\end{equation}
Substituting into (\ref{uneq}) the values of the components of 
$\vec b_1$ and $\vec b_2$ along $\hat C_h$ and $\hat T$, we have that 
\begin{equation} \label{uneq1}
\left\{ \begin{array}{l}
|N_1 n+N_2 m| < N \\
|N_1 t_1+N_2 t_2| < 1 \\
\end{array} \right. \quad .
\end{equation}
Being $N_1 t_1+N_2 t_2$ an integer number, the second inequality is 
equivalent to the relation $N_1 t_1+N_2 t_2=0$. Since $N_1$ and $N_2$ have 
to be integer numbers, and $t_1$ and $t_2$ are two relative prime integer 
numbers, this identity is satisfied only if $N_1=\ell t_2$ and 
$N_2=-\ell t_1$, with $\ell$ an integer number. With these values of $N_1$ 
and $N_2$, we have that $N_1 n+N_2 m=\ell (t_2 n-t_1 m)=-\ell N$
(as we have seen, $m t_1 -n t_2=N$).
Thus the first inequality of (\ref{uneq1}) becomes $|\ell| N < N$ 
and is satisfied only if $\ell=0$. This means that $N_1=\ell t_2=0$
and $N_2=-\ell t_1=0$ and thus $\vec k'$ is identical to $\vec k$, 
as we wanted to prove.


\begin{thebibliography}{999}
\bibitem{meyyappan} Meyyappan~M. Carbon Nanotubes: Science and
Applications. Boca Raton, Florida: CRC Press; 2005.
\bibitem{ajayan} Ajayan~PM, Zhou~OZ. Applications of
Carbon Nanotubes. In: Dresselhaus~MS, Dresselhaus~G, Avouris~Ph, editors. 
Carbon Nanotubes: Synthesis, Structure, Properties, and Applications
(Topics in Applied Physics, vol~80),
Berlin: Springer, 2000; p.~391--425.
\bibitem{hamada} Hamada~N, Sawada~S, Oshiyama~A. New One-dimensional 
Conductors: Graphitic Microtubules.
Phys~Rev~Lett 1992; 68(10), 1579--1581.
\bibitem{mintmire} Mintmire~JW, Dunlap~BI, White~CT.
Are Fullerene Tubules Metallic?.
Phys~Rev~Lett 1992; 68(5), 631--634.
\bibitem{saito1} Saito~R, Fujita~M, Dresselhaus~G, Dresselhaus~MS.
Electronic structure of chiral graphene tubules.
Appl~Phys~Lett 1992; 60(18), 2204--2206.
\bibitem{saito} Saito~R, Dresselhaus~G, Dresselhaus~MS.
Physical Properties of Carbon Nanotubes. 
London: Imperial College Press; 1998.
\bibitem{reichbook} Reich~S, Thomsen~C, Maultzsch~J.
Carbon Nanotubes. Weinheim: Wiley-VCH; 2004.
\bibitem{ashcroft} Ashcroft~NW, Mermin~ND. Solid State Physics. 
London: Brooks/Cole Thomson Learning; 1976.
\bibitem{reich} Reich~S, Maultzsch~J, Thomsen~C, Ordej\'on~P.
Tight-binding description of graphene.
Phys~Rev~B 2002; 66, 035412-1--5.
\end{thebibliography}
\end{document}